\documentclass[letterpaper, 10 pt, conference]{ieeeconf}  % Comment this line out if you need a4paper

\IEEEoverridecommandlockouts                              % This command is only needed if 
\usepackage{color}

\usepackage{cite}
\usepackage{algorithm}
\usepackage{algpseudocode}

\usepackage{amsmath}
\usepackage{multirow}
\usepackage{multicol}   %Columns
\usepackage[justification=centering]{caption}

\usepackage{graphicx}

\usepackage{caption}
\usepackage{subcaption}

\usepackage{epstopdf,float,amsfonts}
\usepackage{epsfig}

\newtheorem{theorem}{Theorem}
\newtheorem{lemma}[theorem]{Lemma}

\newtheorem{assumption}{Assumption}
\let\labelindent\relax 
\usepackage{enumitem}

\usepackage{todonotes} % for comments

\title{\bf \LARGE Neural Lyapunov Differentiable Predictive Control}
\author{Sayak Mukherjee,
           J\'an Drgo\v na, Aaron Tuor, Mahantesh Halappanavar, Draguna Vrabie
\thanks{Authors are with the Pacific Northwest National Laboratory (PNNL), Richland, WA, USA. 
Emails: \{	sayak.mukherjee, jan.drgona, aaron.tuor, mahantesh.halappanavar, draguna.vrabie\}@pnnl.gov.}}
\date{}

\begin{document}

\maketitle

\thispagestyle{empty}
\pagestyle{empty}
\begin{abstract}

We present a learning-based predictive control methodology using the differentiable programming framework
with probabilistic Lyapunov-based stability guarantees. The neural Lyapunov
differentiable predictive control (NLDPC) learns the policy
by constructing a computational graph encompassing the
system dynamics, state and input constraints, and the necessary
Lyapunov certification constraints, and thereafter using the
automatic differentiation to update the neural policy parameters.
In conjunction, our approach jointly learns a Lyapunov function
that certifies the regions of state-space with stable dynamics.
We also provide a sampling-based statistical
guarantee for the training of NLDPC from the distribution of initial conditions. Our offline training approach provides a computationally efficient and scalable alternative to classical explicit model predictive control solutions. We substantiate the advantages of the proposed approach with simulations to stabilize the double integrator model and on an example of controlling an aircraft model.
\end{abstract}

 \begin{keywords} 
 Differentiable predictive control, model predictive control, neural policy, Lyapunov guarantee.
\end{keywords}

\section{Introduction}

% AIM: CDC22, deadline March 30th

% CDC: focus on linear systems and interplay between lyapunov condition and constraints

% journal extension: Offline Learning of Constrained Neural Lyapunov Differentiable Predictive Control Policies for Nonlinear Systems
 The optimization-based solution to the model predictive control (MPC) problems can become restrictive in some real-life applications due to the computational burden of computing implicit control policies by solving nonlinear optimization programs. Explicit MPC (eMPC) provides an alternative solution at the control inputs are pre-computed to avoid online optimization for certain prespecified parameters~\cite{BemEtal:aut:02,Alessio2009}. However, the multiparametric optimization problem quickly becomes intractable for larger systems, thereby limiting its practical usability. 
Therefore, in the last decade, researchers have focused on reducing the computational complexity of the eMPC problems~\cite{KVASNICA20131776,HOVLAND20087711}.
More recent trends show promises in developing data-driven predictive control solutions for the underlying explicit MPC problem. 

Learning-based MPC (LBMPC)~\cite{ASWANI20131216} methods utilizes the learned system dynamic model from trajectory measurements via adaptive MPC frameworks~\cite{aswani2014practical,Hewing2020}. Learning-based approaches are also supplemented with various forms of robustness conditions and capture state-dependent uncertainties 
with work such as~\cite{SOLOPERTO2018442}, with guarantees on bounded uncertainties in~\cite{Bujarbaruah2018AdaptiveMF}, to name a few.
On the other hand, approximate MPC~\cite{Domahidi2011, ZhangKLA15,DRGONA2018, RahaCRB20,chen2019large}
utilizes a supervised learning framework to imitate the  MPC trajectories, which, although efficient in implementation, is bottlenecked by its inherent dependency on the original MPC problem solutions.
From the computer science perspective, researchers have focused on developing new data-driven model architectures inspired by control theory.
Work such as~\cite{MPC_PInet2018}  have explored MPC-inspired neural architectures for learning neural control policies.
The use of convex neural networks~\cite{AmosXK16} for optimal predictive control problems is investigated in~\cite{chen2018optimal}. 
While~\cite{DiffCVxLayers2019} introduced implicit layers to solve constrained optimization problems within deep neural networks leading to works such as~\cite{GNURL2019,maddalena2019neural}. 
Others~\cite{diffMPC2018,East2020} have proposed the use of automatic differentiation to compute the sensitivities of the MPC problems for safe control in the context of imitation learning.
To provide guarantees,
~\cite{Donti_control2020} leverage linear matrix inequality (LMI) framework for enforcing Lyapunov stability conditions while learning neural control policies.
A semidefinite programming-based safety verification and robustness of neural network control policies have been investigated in~\cite{Mahyar2022}. \cite{LyapunovNN2018} presents adaptive safe learning with Lyapunov provable safety certificate. \cite{chang2019neural} presents a learner-falsifier-based Lyapunov guarantee for learning neural policies for optimal control of known dynamics.
An alternative approach presented by~\cite{Zanon2019,Gros2021} provides safety guarantees for reinforcement learning (RL) algorithms by representing 
 the control policy by an implicit differentiable MPC. \cite{mittal2020neural} proposes to use neural Lyapunov functions to tune parameters of online implicit MPC. \cite{saha2021neural} presents a system identification methodology for nonlinear dynamics to jointly learn a controller
and corresponding stable closed-loop dynamics.

Inspired by these trends, in~\cite{drgona2022learning} we have introduced the differentiable predictive control (DPC)
framework as a scalable data-driven alternative to analytical multiparametric programming solvers.
DPC brings forth the idea of offline computation of explicit predictive control policies by leveraging automatic differentiation (AD) of the constrained optimization problem for direct computation of the policy gradients. 
In our previous work, we have demonstrated the scalability of the DPC framework on systems, including uncertainties and nonlinear constraints~\cite{drgovna2022learning_spdpc,DRGONA202114}.

\noindent \textbf{Contributions.}
This paper brings forth the idea of jointly learning the Lyapunov function candidates to provide guarantees in the DPC policy learning framework. 
We parametrize the Lyapunov function using deep neural network architecture introduced in~\cite{NIPS2019_9292}. Then we solve the parametric optimal control problem with such Lyapunov certification conditions as a constraint via the DPC policy optimization algorithm~\cite{drgona2022learning}. The differentiable programming framework constructs a computational graph of the closed-loop parametrized system with neural predictive policy and learnable Lyapunov functions and then uses AD to backpropagate resultant loss using MPC objectives and soft constraint penalties to update policy and Lyapunov network parameters. Additionally, we provide a theoretical framework for guaranteeing stability and constraint satisfaction during the training, where we sample initial conditions from a given distribution and characterize the system trajectories based on such initial conditions in a probabilistic manner. For such analysis, we utilize Hoeffding's inequality adapting from \cite{hertneck2018learning} which was described in the context of unsupervised learning of the constrained control policies. We perform detailed numerical experiments on stabilizing the double integrator model and controlling an aircraft model to bring out various intricacies of our design. 

The paper is organized as follows. The problem formulation and necessary background is discussed in Section II. The proposed method is described in Section III with the necessary building blocks and algorithm. Section IV describes the probabilistic guarantees ensuring stability and constraint satisfaction with sampled initial conditions. Numerical experiments are presented in detail in Section V, and we provide concluding remarks in Section VI.

% \add{problem = simultaneous learning of constrained predictive control neural policies + lyapunov functions}

% \add{define constrained closed loop dynamical system in discrete time, state what is known (system model and constraints) and what is unknown (policy) }

\section{Problem Formulation and Background}

We consider the discrete-time non-linear dynamic system:
\begin{align}
\label{discrete_dynamics}
    {\bf x}_{k+1} = f( {\bf x}_k, {\bf u}_k), \  k \in \mathbb{N}_{0}^{N-1},
\end{align}
where ${\bf x}_k \in \mathbb{R}^{n_x}$ is the system state, ${\bf u_k} \in \mathbb{R}^{n_u}$ is the control input at time $k$, and $\mathbb{N}_a^b = \{a, a+1, \ldots, b \}$ is a set of integers. We assume the dynamics $f(.)$ is locally Lipschitz continuous and known. The system is also constrained by state and control input based constraints.

\subsection{Parametric Optimal Control Problem}

% \add{lets rephrame this problem as parametric optimal control problem (explicit MPC) not to create confusion that we solve online MPC. Lets do not re-use the previous text from our old paper, but write everything from scratch so we don't have the same structure repeated. }

% \add{Jan: lets add all background on DPC, including objectives, penalties, neural policies, and assumptions on differentiability here}

 We intend to 
 solve the following parametric optimal control problem:
 \begin{subequations}
\label{eq:POCP}
    \begin{align}
 \min_{\boldsymbol \theta} \ & \sum_{k=0}^{N-1}\ell( {\bf x}_k, {\bf u}_k), \label{eq:mpc_example:Q} & 
 \\ 
  \text{s.t.} \ &   {\bf x}_{k+1} =  f( {\bf x}_k, {\bf u}_k), \  k \in \mathbb{N}_{0}^{N-1},     & \\
 \  &  [{\bf u}_0, \ldots, {\bf u}_{N-1} ] = \boldsymbol{\pi}_{ \theta}({\bf x}_0), \label{eq:mpc_example:u}\\
  &  h({\bf x}_k) \le {\bf 0},
 \\
 &   g({\bf u}_k) \le {\bf 0}, \\
   &  {\bf x}_0 \sim P_{{\bf x}_0}. \label{eq:mpc_example:init}
\end{align}
\end{subequations}
 where ${\bf x}_k  \in \mathbb{R}^{ n_{x}}$,  ${\bf u}_k  \in \mathbb{R}^{ n_{u}}$ are states and control actions, respectively,
 $N$ defines finite time prediction horizon, and $\boldsymbol \theta$ are the parameters of the control policy~\eqref{eq:mpc_example:u} to be optimized over a distribution of initial conditions~\eqref{eq:mpc_example:init}.
 
We consider the objective~\eqref{eq:mpc_example:Q} to be a differentiable function. We can consider quadratic objective functions with state and control energy minimization such as:
\begin{equation}
\label{eq:MPC_obj}
     \ell( {\bf x}_k, {\bf u}_k, {\bf r}_k  ) =
 || {\bf x}_k ||_{Q_x}^2  + || {\bf u}_k ||_{Q_u}^2 
\end{equation}
with  $\|{\bf a}\|_Q^2 = {\bf a}^T Q {\bf a}$ being weighted squared $2$-norm. We consider the following assumptions:

\begin{assumption}
The nominal system dynamics model~\eqref{discrete_dynamics} is controllable, and known. The dynamics $f(.)$ can be a learned model or a physics based model.
\end{assumption}

\begin{assumption}
The parametric optimal control objective $  \ell( {\bf x}_k, {\bf u}_k)$
and constraints $  h({\bf x}_k)$, and $  g({\bf u}_k)$, respectively, are at least once differentiable.
\end{assumption}

\subsection{Neural Predictive Control Policy}

% \add{Jan: we can have a short paragraph on assumptions about our policy structure}

Without the specific parametric policy structure as in~\eqref{eq:mpc_example:u}, then the classical MPC solution provides an implicit solution for the control inputs utilizing nonlinear programming based solvers. On the contrary, we are interested in generating explicit model predictive (eMPC) control policies with a specific parametric control structure. For the system dynamic model~\eqref{discrete_dynamics}, we want to compute parametric predictive control policies modeled by deep neural networks 
$ \pi_{\boldsymbol \theta}({\bf x}_0): \mathbb{R}^{n_x} \to \mathbb{R}^{N \times n_u} $~\eqref{eq:dnn} that minimizes the parametric control objective function~\eqref{eq:mpc_example:Q} for prediction window of  $N$ steps, while satisfying the constraints. 
We consider the policy to be made of fully connected neural networks such as:
\begin{subequations}
    \label{eq:dnn}
    \begin{align}
   {\bf U} = \boldsymbol{\pi}_{ \boldsymbol \theta}({\bf x}_0) & =  \mathbf{W}_{L}  \mathbf{z}_L + \mathbf{b}_{L}, \\
    \mathbf{z}_{l} &= \boldsymbol\sigma(\mathbf{W}_{l-1} \mathbf{z}_{l-1} + \mathbf{b}_{l-1})  \label{eq:dnn:layer},\\
    \mathbf{z}_0 &=  {\bf x}_0,
 \end{align}
\end{subequations}
where ${\bf U} = [{\bf u}_0, \ldots, {\bf u}_{N-1} ] \in \mathbb{R}^{N \times n_{u}}$
represents the generated control sequence over $N$ time steps. 
Vector ${\bf x}_0$ represents full state initial condition feedback, and subsequently the vector ${\bf x}_k$ provides the instantaneous state vector at time step $k$.
Considering the architecture of the policy class,
 $\mathbf{z}_i$ represents hidden states, $\mathbf{W}_i$, and $\mathbf{b}_i$  are the weights and biases of the $i$-th layer, respectively. These policy parameters are compactly represented as $\boldsymbol \theta$ which we intend to optimize.
The nonlinearity $\boldsymbol\sigma: \mathbb{R}^{n_{z_l}} \rightarrow \mathbb{R}^{n_{z_l}}$ is constructed by executing differentiable activation function $\sigma: \mathbb{R} \rightarrow \mathbb{R}$ in an element-wise manner.

\subsection{Stability via Neural Lyapunov Functions}

We intend to provide Lyapunov stability certificate during the training of the explicit neural predictive control policy $\pi_{\boldsymbol \theta}({\bf x}_0)$. The closed-loop control at time step $k$ can be computed by taking the first control action of the predictive sequence denoted by ${\bf u}_k = [\pi_{\boldsymbol \theta}({\bf x}_k)]_0$, however with slight abuse of notation we continue to use the notation $\pi_{\boldsymbol \theta}({\bf x}_k)$ as our control input ${\bf u}_k$. The Lyapunov function $V(\bf x_k)$ can characterize the stability of the closed-loop dynamics as follows.
\begin{lemma}
(Lyapunov Characterization) : Using the neural predictive control policy ${\bf u}_k = \pi_{\boldsymbol \theta}({\bf x}_k)$ for discrete time steps $k$,  with the Liptsitz continuous closed-loop $f(\bf x_k, {\bf u}_k)$, for the equilibrium $\bf x_\mathcal{O} = \bf 0$, if there exists a set $\mathcal{D} \subset \mathbb{R}^n$ on which the function $V(.)$ is always positive-definite, and satisfy,
\begin{align}
    \label{Discrete-time Lyap}
    V(f({\bf x}_k, {\bf u}_k)) - V({\bf x}_k) < 0, \forall {\bf x}_k \in \mathcal{D} - \mathbf{0},
\end{align}
then $\bf x_\mathcal{O} = \bf 0$ is an asymptotically stable equilibrium, and $V(.)$ is a control Lyapunov function (CLF) for the closed-loop dynamics.

\end{lemma}

We consider a parametrized closed-loop Lyapunov function, the CLF, with deep neural network parameters $\phi$, denoted as $V_\phi ({\bf x}_k)$. As proposed in \cite{kolter2019learning}, we can consider the following structure for the Lyapunov function candidate based on input-convex neural network (ICNN):
    \begin{align}
    \label{ICNN}
&{\bf z}_{1} =\sigma_{0}\left({\bf W}_{0} {\bf x}+{\bf b}_{0}\right), \\
&{\bf z}_{i+1} =\sigma_{i}\left({\bf U}_{i} {\bf z}_{i}+{\bf W}_{i} {\bf x}+{\bf b}_{i}\right), i=1, \ldots, k-1, \\
&\bf g(x) \equiv {\bf z}_{k},
\end{align}
which makes the function $\bf g(x)$ convex in $\bf x$.
The activations $\sigma_k(.)$ are are
monotonically non decreasing convex nonlinear scalar functions, such as the ReLU, $\bf U_i$ are positive weight mappings, and other weights and biases are real-valued. To make the candidate Lyapunov function positive definite another compositional operation is performed as:
\begin{align}
    V({\bf x})=\sigma_{k+1}({\bf g(x)-g(0)})+\epsilon\| {\bf x} \|_{2}^{2},
\end{align}
where $\sigma_{k}(.)$ is a positive convex non-decreasing function with $\sigma_{k}(0) = 0$ and $\epsilon $ is a small constant. To make the Lyapunov function continuously differentiable we can consider smooth \texttt{ReLU} activation as mentioned in~\cite{kolter2019learning}. 

At this point, we state our problem statement as: \\
\noindent \textbf{P.} \textit{Learn the neural predictive control policy $\bf u_k = \pi_{\boldsymbol \theta}({\bf x}_k)$ for the parametric optimal control problem~\eqref{eq:POCP} along with simultaneous Lyapunov certification using the parametric control Lyapunov function $V_\phi ({\bf x}_k)$ following Lemma 1 such that the  closed-loop $f(\bf x_k, \bf u_k)$  dynamics remain Lyapunov-stable along with solving for the desired optimal control performance.}

% \add{now we have all the basic building block to construct our Lyapunov DPC}

%\add{This whole new section needs to be rewritten not just tweeked}

\section{Neural Lyapunov Differentiable Predictive Control (NLDPC) Framework}

In differentiable predictive control, we train the explicit predictive control policy by utilizing the system dynamics $f(.)$ to generate predictive trajectories in a self-supervised way. We assume that the system's state dynamics can be obtained by performing system identification or mathematical modeling of the underlying physics.
The system dynamics and the neural predictive control architecture are combined in a computational graph to create a differentiable closed-loop representation.
The parameters of the control policy are updated via stochastic gradients descent by backpropagating the gradients of the optimal control objective function and constraints penalties. 
In this work, we expand the DPC framework to allow for joint learning of  neural predictive control
policies and neural Lyapunov functions enforcing closed-loop stability of the controlled system.
We first discuss the penalty-based approach in dealing with the state, control constraints, and the Lyapunov-based certification criterion.

\subsection{Soft state, control, and Lyapunov constraints }

% The DPC problem~\eqref{eq:DPC}  is  solved by differentiating the parametric MPC loss in the Lagrangian form. Here, the first term of the objective gives a main performance metric, e.g., reference tracking loss
% $\ell_{\texttt{MPC}}( {\bf x}_k, {\bf u}_k, {\bf r}_k ) =
%  || {\bf x}_k - {\bf r}_k ||_{Q_r}^2  + || {\bf u}_k ||_{Q_u}^2 $ parametrized by ${\bf r}_k$.
The state and action constraints $h({\bf x}_k), g({\bf x}_k)$, and the control Lyapunov certification criterion $V_\phi ({\bf x}_k)$  as shown in Lemma 1 are modeled as soft constraints using the \texttt{ReLU} activation function represented by $p_x({\bf x}_k)$, $p_u({\bf u}_k)$, and $p_V({\bf x}_{k+1}, {\bf x}_k)$,  respectively as follows:
\begin{subequations}
\label{eq:ReLU_ineq}
    \begin{align}
    p_x({\bf x}_k) & =  ||
    \texttt{ReLU}(h({\bf x}_k)||_2 \\
p_u({\bf u}_k)  & =  || \texttt{ReLU}(g({\bf u}_k))||_2 \\
 p_V({\bf x}_{k+1}, {\bf x}_k) & =   || \texttt{ReLU}(V_{\phi}({\bf x}_{k+1}) - V_{\phi}({\bf x_k}) )||_2,
 \end{align}
 \end{subequations}
% \begin{subequations}
% \label{eq:ReLU_ineq}
%     \begin{align}
%     p_x(h({\bf x}_k)) & =
%     \frac{1}{n_h}  \sum_{j=1}^{n_h} 
%     \texttt{ReLU}(h_j({\bf x}_k)), \\
% p_u(g({\bf x}_k) )  & = \frac{1}{n_g}  \sum_{j=1}^{n_g}  \texttt{ReLU}(g_j({\bf x}_k)),
%  \end{align}
%  \end{subequations}
 with $\texttt{ReLU}$ standing for rectifier linear unit function. This structural representation will help us to formulate the neural Lyapunov DPC problem as described in the next sub-section. 

%  $n_h$ and  $n_g$ defining the total number of state and input constraints, and $l$  representing the penalty norm. 

\subsection{Neural Lyapunov DPC Problem}

To formulate the learning-based predictive control problem, we utilize the batched training strategy where we generate multiple batches of randomly sampled initial condition that will be used for forward pass roll outs of the DPC computational graph. 
We sample the initial conditions following user-defined distribution $\mathcal{D}$:
\begin{align}
    {\bf x}_0 \sim \mathcal{D}, \ {\bf x}_0 \in \mathbb{X} \subseteq \mathbb{R}^{n_{x}}.
\end{align}

Putting all together, we formulate the neural Lyapunov DPC as following sampled parametric optimal control problem:
\begin{subequations}
\label{eq:DPC}
    \begin{align}
     \label{eq:DPC:objective}
 \min_{{\bf \theta, \phi}} & \frac{1}{mN} \sum_{i=1}^{m} \sum_{k=0}^{N-1}  \big( \ell( {\bf x}_k^i, {\bf u}_k^i ) + Q_V p_V({\bf x}_{k+1}^i, {\bf x}_k^i) +   & \\  & Q_x p_x({\bf x}_k^i)  + Q_u p_u({\bf u}_k^i)  \big),& 
 \nonumber \\ 
  \text{s.t.} \ &   {\bf x}_{k+1}^i = f( {\bf x}_k^i, {\bf u}_k^i), \  k \in \mathbb{N}_{0}^{N-1} ,  \label{eq:dpc:x}  & \\
 \  & [{\bf u}_0, \ldots, {\bf u}_{N-1} ]^i = \boldsymbol \pi_{ {\bf \theta}}({\bf x}_k^i)  \label{eq:dpc:pi}, \\
      \ &  {\bf x}_0^i \in \mathbb{X} \subseteq \mathbb{R}^{n_{x}} \sim \mathcal{D} \label{eq:dpc:x0}.
\end{align}
\end{subequations}
here the index $i$ denotes the batch sample. The  DPC loss function is made of multiple terms, i.e., the optimal control objectibe $ \ell( {\bf x}_k, {\bf u}_k): \mathbb{R}^{n_x + n_u } \to \mathbb{R}  $,
and penalties of parametric constraints  $p_x(h({\bf x}_k)): \mathbb{R}^{n_h } \to \mathbb{R}   $,
$ p_u(g({\bf u}_k)): \mathbb{R}^{n_g} \to \mathbb{R} $, and $p_V({\bf x}_{k+1}, {\bf x}_k): \mathbb{R} \to \mathbb{R}   $. $Q_V, Q_x$ and $Q_u$ represent the scalar penalty weights.

 To utilize differentiable programming, we keep the following assumption.
 \begin{assumption}
\label{assume:loss_con_diff}
The optimal parametric objective function $\ell( {\bf x}_k, {\bf u}_k)$, and state, input and Lyapunov constraint penalties $p_x({\bf x}_k)$, $p_u({\bf u}_k)$, and $p_V({\bf x}_{k+1}, {\bf x}_k)$ are differentiable  at almost every point in their domain.
\end{assumption}

% \add{we need new graphical methodology here, overall problem formulation of DPC with Lyapunov constraints }

\subsection{Neural Lyapunov DPC Computational Graph}

The formulation~\eqref{eq:DPC} is being implemented in the form of a differentiable program for optimizing the neural parameters $\theta$ and $\phi$ for neural predictive policy and Lyapunov function respectively. We construct the computation graph for the differentiable program which optimizes the neural parameters
by differentiating the neural Lyapunov DPC loss function~\eqref{eq:DPC:objective} and backpropagating the gradients through the parametrized closed-loop dynamics given by the system model~\eqref{eq:dpc:x}  and neural control policy~\eqref{eq:dpc:pi}.
\begin{figure*}[htb!]
    \centering
    \includegraphics[width = \linewidth]{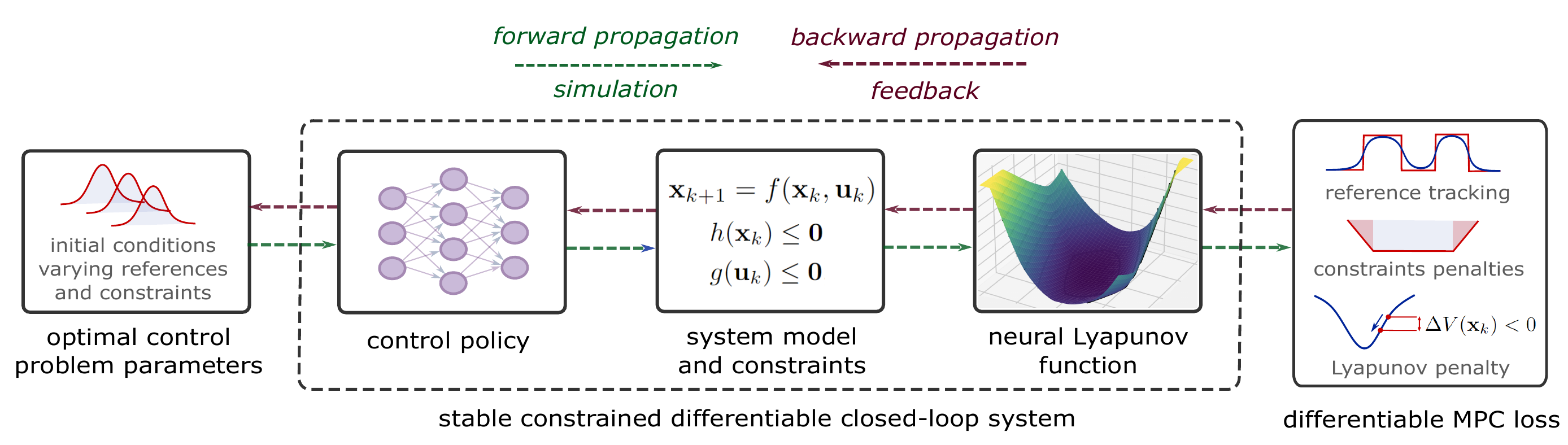}
    \caption{Neural Lyapunov DPC methodology.}
    \label{fig:DPC_graph}
\end{figure*}
% 
% \begin{figure}
%     \centering
%     \includegraphics[width = \linewidth]{./figs/problem_graph.png}
%     \caption{Neural Lyapunov DPC computational graph.}
%     \label{fig:DPC_graph}
% \end{figure}
Fig.~\ref{fig:DPC_graph} shows the data flow in the computational graph of the presented neural Lyapunov DPC method. We have developed a generic framework of output feedback structure where we can use an estimator parametrized by deep neural networks to estimate the initial condition from a past output and input trajectories. However, for the simplicity of the exposition in this paper, we consider a full state feedback. The policy component reads the current states and generates predictive control trajectories over the prediction horizon. Thereafter the initial state and control  trajectories are used to generate a $N$-step ahead rollout of the system dynamics obtaining state trajectories.
The generated trajectories are then passed through penalty-based constraints including Lyapunov constraint~\eqref{Discrete-time Lyap}.

\subsection{Neural Lyapunov DPC Gradients}

Finally, all these individual components described above lead to the neural NLDPC loss function $\mathcal{L}_{\texttt{NLDPC}}$~\eqref{eq:DPC:objective}. 
One can use stochastic gradient descent and its variants to optimize the loss function  $\mathcal{L}_{\texttt{NLDPC}}$ over the sampled distribution of initial conditions~\eqref{eq:dpc:x0}. 
The differentiable closed-loop dynamics with the learnable policy, optimal control objective, and constraints allow us to use automatic differentiation \cite{puskorius1994truncated} to directly compute policy gradients. 
As described before, the problem~\eqref{eq:DPC} is represented as a computational graph, and therefore, using chain rule, one can compute the gradients of $\mathcal{L}_{\texttt{NLDPC}}$ w.r.t. the policy parameters $\theta$ as follows:
\begin{equation}
\label{eq:grad}
\begin{split}
    \nabla_{\theta} \mathcal{L}_{\texttt{NLDPC}} =  \frac{ \partial\ell( {\bf x}, {\bf u} )}{\partial \theta} + \frac{ \partial p_V({\bf x})}{\partial \theta} + \\ \frac{ \partial p_x({\bf x})}{\partial \theta} + 
    \frac{ \partial p_u({\bf u})}{\partial \theta} = \\
    \frac{ \partial\ell( {\bf x}, {\bf u})}{\partial {\bf x}}   \frac{ \partial {\bf x}}{\partial {\bf u}} \frac{ \partial {\bf u}}{\partial \theta} +   \frac{ \partial\ell( {\bf x}, {\bf u})}{\partial {\bf u}}  \frac{ \partial {\bf u}}{\partial \theta} +  \frac{ \partial p_V({\bf x})}{\partial {\bf x}}  \frac{ \partial {\bf x}}{\partial {\bf u}}\frac{ \partial {\bf u}}{\partial \theta} \\
    \frac{ \partial p_x({\bf x})}{\partial {\bf x}}  \frac{ \partial {\bf x}}{\partial {\bf u}} \frac{ \partial {\bf u}}{\partial \theta} +   \frac{ \partial p_u({\bf u})}{\partial {\bf u}} \frac{ \partial {\bf u}}{\partial \theta},
\end{split}
\end{equation}
where $\frac{ \partial {\bf u}}{\partial \theta}$ represents partial derivatives of the neural policy with respect to its weights, computed using backpropagation. Similarly, we can compute $ \nabla_{\phi} \mathcal{L}_{\texttt{NLDPC}}$, i.e., gradients with respect to neural Lyapunov function parameters. The
parametric optimal control problem~\eqref{eq:DPC} now can be solved by using scalable stochastic gradient optimization algorithms such as AdamW~\cite{loshchilov2017decoupled}. This provides us with a way to optimize the policy offline.
In practice, we use automatic differentiation frameworks such as Pytorch~\cite{paszke2019pytorch} to compute these gradients in the NLDPC computational graph.

% show the generated graph, and explain how it can be used to compute sensitivities in the follow up section
% The differentiable predictive control architecture is built upon the computational graph encompassing the dynamic system and neural control policy such that we can formulate differentiable loss functions that consider constraints as soft penalties. 

% \subsection{DPC Gradients}

% derive backprop sensitivity of the parametric optimal control problem via chain rule

% todo: sensitivities of the policy parameters and lyapunov parameters separately

% define DPC loss as eq 7a and 7b

\subsection{Simultaneous Learning of Constrained Neural Control Policy and  Lyapunov Function }

% \add{ here we add derived policy and Lyapunov function gradients, and maybe adding figure of the resulting computational graph}

% \add{here we define the offline learning algorithm using the formulation and gradients from previous subsections. We don't do any hypothesis testing or falsification here }

We now describe how to learn both a control policy and a neural Lyapunov function together, so
that the Lyapunov conditions can be rigorously verified to ensure the stability of the system. $ \nabla_{\theta} \mathcal{L}_{\texttt{NLDPC}}$, and $ \nabla_{\phi} \mathcal{L}_{\texttt{NLDPC}}$ help us to train these policies and neural Lyapunov function simultaneously with batches of state trajectory data for different samples of initial conditions. The NLDPC computational graph has been used along with reverse-mode automatic differentiation framework to implement the algorithm. Algorithm $1$ presents the steps of the proposed framework.

\begin{algorithm}[ht!]
  \caption{Neural Lyapunov DPC (NLDPC)}\label{algo:NLDPC}
  \begin{algorithmic}[1]
  \State \textbf{Input} identified or physics based system dynamics $f(.)$
 %\State \textbf{function}   
  \hspace{1cm} 
  \State \textbf{Input} candidate Lyapunov function $V_\phi (\bf x)$
    \State \textbf{Input} optimal control loss $\ell(\bf x, u)$ and constraints $  h({\bf x})$, and $  g({\bf u})$
 \State \textbf{Input} optimizer $\mathbb{O}$
 \State \textbf{Construct} NLDPC computational graph based on \eqref{eq:DPC} 

 \For{epochs=1:$N_{\texttt{epoch}}$} 
     \State \textbf{Sample} initial conditions ${\bf x}_0$ from distribution $\mathcal{D}$
    \State \textbf{Evaluate} forward pass of the NLDPC computational 
    \State graph and compute  $\mathcal{L}_{\texttt{NLDPC}}$
    \State \textbf{Differentiate} NLDPC computational graph  w.r.t. 
    \State policy and Lyapunov candidate parameters to obtain 
    \State gradients $\nabla_{\theta} \mathcal{L}_{\texttt{NLDPC}}$ and $\nabla_{\phi} \mathcal{L}_{\texttt{NLDPC}}$
%   $
% V_{\theta}(x), u(x) \leftarrow \mathrm{NN}_{\theta, u}(x)$
% \State 
% $\nabla_{f_{u}} V_{\theta}(x) \leftarrow \sum_{i=1}^{D_{i n}} \frac{\partial V}{\partial x_{i}}\left[f_{u}\right]_{i}(x) $
% \State 
% $\text { Compute Lyapunov risk } L(\theta, u) $
% \State Compute Lyapunov DPC loss  $\mathcal{L}_{\texttt{L-DPC}} = L(\theta, u)  + \mathcal{L}_{\texttt{DPC}}$
\State \textbf{Update}  control parameters
$\theta \leftarrow \theta +\alpha \nabla_{\theta} \mathcal{L}_{\texttt{NLDPC}}
$
\State \textbf{Update} Lyapunov parameters 
$\phi \leftarrow \phi+\alpha \nabla_{\phi} \mathcal{L}_{\texttt{NLDPC}}$ 
\EndFor
\State \textbf{Return}: trained policy $\boldsymbol \pi_{ \theta}$ and Lyapunov function $V_{\phi}$
  \end{algorithmic}
\end{algorithm}

\section{Stability Guarantees}

Probabilistic considerations have been discussed in works \cite{hertneck2018learning, rosolia2018stochastic, Karg2021} for learning-based MPC setting. We use sampling-based guarantees for the neural Lyapunov DPC framework extending the approach of \cite{hertneck2018learning}. The parametric optimization problem~\eqref{eq:DPC} has been solved using Algorithm \ref{algo:NLDPC} where we created multiple batches with
sampled initial conditions ${\bf x}_0^i \in \mathbb{X} \subseteq \mathbb{R}^{n_{x}}$ from a distribution $\mathcal{D}$. The variations in the initial condition are shown with superscript $i$, and we create $m$ such scenarios.

% However, theoretically it is not straightforward to consider such amount of uncertainties to provide feasibility and stability considerations for the closed-loop. Therefore, we supplement the learning problem with few specified considerations to provide probabilistic guarantees. First we add the terminal constraint on the states such that,
% \begin{align}
% \label{eq:terminal}
%     {\bf x}_{N} \in  \mathcal{X}_f.
% \end{align}
% Under uncertainties due to external disturbances, system parameters and initial conditions, the solution trajectories from \eqref{eq:dpc:x} may violate state and parameter dependent constraints. Therefore, in this paper, we adopt chance constraint  framework~\eqref{eq:chance_con}.

We denote the set of closed-loop state rollouts by  ${\bf \mathcal{T}}^{i}$, and the
control trajectories provided by the learned NLDPC policy by $[{\bf u}_0^{i}, \ldots, {\bf u}_{N-1}^{i}] = \boldsymbol \pi_{ \theta}({\bf x}_0^{i})$,
for  $\ i \in \mathbb{N}_{1}^{m}$, i.e.,
\begin{align}
    {\bf \mathcal{T}}^{i}: \begin{cases} &\boldsymbol \{ {\bf x}_k^{i}\}, \ \forall k \in \mathbb{N}_{0}^{N-1},   \nonumber\\
    &{\bf x}_{k+1}^{i} = {\bf A} {\bf x}_k^{i} + {\bf B } {\bf u}_k^{i}.
    \end{cases}
\end{align}
Considering the constraints, we can create a probabilistic metric for the sampled roll-outs such as,
\begin{align}
    {\bf P}^{i}= \mbox{True} : \begin{cases}
   h({\bf x}_k^{i}) \le {\bf 0}, \ \forall k \in \mathbb{N}_{0}^{N-1},
 \\
  g({\bf u}_k^{i}) \le {\bf 0}, \ \forall k \in \mathbb{N}_{0}^{N-1}, \\
  V_\phi (f({\bf x}_k^i, {\bf u}_k^i)) - V({\bf x}_k^i) < 0,
    \end{cases}
\end{align}
along with the indicator:
\begin{align}
    \mathcal{I}({\bf \mathcal{T}}^{i}) := \begin{cases}
    \label{eq:Indicator}
    1 \;\; \text{if} \;\; {\bf P}^{i}=\mbox{True}.\\
    0, \;\; \text{otherwise},
    \end{cases}
\end{align}
which indicates if the learning based policy $\boldsymbol \pi_{ \theta}({\bf x}_0^{i})$ satisfies the required constrained during the sampled roll-out.

\begin{theorem}
 Consider sampled initial conditions for solving the parametric optimal control problem~\eqref{eq:DPC} with assumptions $1-3$, and the predictive control policy using Algorithm~\ref{algo:NLDPC}. Selecting constraint violation probability $1-\kappa$, and level of confidence parameter $\delta$, if the empirical risk  on the indicator function \eqref{eq:Indicator}, denoted as $\tilde{\sigma}$, with sufficiently large number of sample trajectories $m$ satisfy $\kappa \le
    \tilde{\sigma} - \sqrt{-\frac{\ln{\frac{\delta}{2}}}{2m}}$, then  the learned NLDPC policy $\boldsymbol\pi_{ \theta}({\bf x}_k)$ will constraint satisfaction and closed-loop stability in probabilistic sense.  
\end{theorem}

\textit{Proof:} We generate $m$ number of sampled initial conditions which lead to $m$ number of trajectories. The initial conditions are sampled in an i.i.d. manner, resulting in i.i.d. ${\bf \mathcal{T}}^{i}, \mbox{and} \;\; \mathcal{I}({\bf \mathcal{T}}^{i})$. The empirical risk for all the trajectories is given by,
\begin{align}
    \Tilde{\sigma} = \frac{1}{m}\sum_{i=1}^m \mathcal{I}({\bf \mathcal{T}}^{i}).
\end{align}

We can guarantee stability and constraint satisfaction if for ${\bf x}_0^{i} \sim \mathcal{D}, 
  $ we have, $\mathcal{I}({\bf \mathcal{T}}^{i}) = 1$, i.e. $h({\bf x}_k^{i}) \le {\bf 0}$,
  $  g({\bf x}_k^{i}) \le {\bf 0}$, $V_\phi (f({\bf x}_k^i, {\bf u}_k^i)) - V({\bf x}_k^i) < 0$, ${\bf x}_N^{i} \in \mathcal{X}_T$, $\forall i \in \mathbb{N}_1^m$.
   
   Let us denote $\sigma := \textbf{Pr}(\mathcal{I}({\bf \mathcal{T}}^{i}) = 1)$. Recalling Hoeffding’s Inequality \cite{hertneck2018learning} to estimate $\sigma$ from $\tilde{\sigma}$ we can write:
%\textit{Lemma 1} [Hoeffding’s Inequality]:    
% Let $\mathcal{I}(\bf X^{i})$ be $r$ iid random variables with $0 \leq \mathcal{I}(\bf X^{i}) \leq 1$. One can have,
\begin{align}
    \textbf{Pr}(|\tilde{\sigma} - \sigma| \geq \alpha) \leq 2\mbox{exp}(-2r\alpha^2) \;\; \forall \alpha > 0.
\end{align}

Thereafter, using $\delta := 2\mbox{exp}(-2r\alpha^2)$, with confidence $1 - \delta$ one can have,
\begin{align}
\label{eq:prob_guarantee}
   \textbf{Pr}(\mathcal{I}({\bf \mathcal{T}}^{i}) = 1) = \sigma \geq \tilde{\sigma} - \alpha.
\end{align}

Therefore with a chosen confidence $\delta$ and risk lower bound $\kappa \leq \textbf{Pr}(\mathcal{I}({\bf \mathcal{T}}^{i})) = 1$, the empirical risk can be bounded as:
\begin{equation}
\label{eq:mu_bound}
   \kappa \le \tilde{\sigma} - \alpha =
    \tilde{\sigma} - \sqrt{-\frac{\ln{\frac{\delta}{2}}}{2m}}
\end{equation}
In a nutshell, once we fix the confidence $\delta$ and risk lower bound $\kappa$, the empirical risk $\tilde{\sigma}$ and $\alpha$ is evaluated for an experimental value of $m$. Therefore when policies trained via Algorithm~\ref{algo:NLDPC} satisfies \eqref{eq:mu_bound}, then with confidence at least $1-\delta$, at least a fraction of $\kappa$ trajectories ${\bf \mathcal{T}}^{i}$ will satisfy ${\bf P}^{i}=\mbox{True}$. Therefore, state, action and Lyapunov constraints are  satisfied with the confidence $1-\delta$.
% Furthermore, along with the constraint satisfaction of the closed-loop trajectories, assumption $4$ guides us to the existence of a positive invariant terminal set in presence of bounded disturbances, thereby maintaining stability once the constraints are satisfied in probabilistic sense. 
This concludes the proof. 
% guaranteeing  stability and constraint satisfaction of the learned policy using the SP-DPC optimization algorithm  Algorithm~\ref{algo:DPC_optim}.  \qed

\section{Numerical Case Studies}

% \add{we do 2-3 linear systems subject to constraints: double integrator, pvtol aricraft model, maybe double integrator with obstacle avoidance or quadcopter model}
\subsection{Example 1: Stabilizing Double Integrator}
We start with the task of learning a stabilizing neural feedback policy for an unstable system. The unstable double integrator model is given as: 
 \begin{align}
 \label{eq:double_int}
{\bf x}_{k+1} = \begin{bmatrix} 1.2 & 1.0 \\ 0.0 & 1.0\end{bmatrix}  {\bf x}_k + \begin{bmatrix} 1.0  \\ 0.5  \end{bmatrix}   {\bf u}_k.
 \end{align}
The system is subject to state and input constraints given as 
 ${\bf x}_k \in \mathcal{X} := \{ {\bf x} \mid -10 \le { x} \le 10 \}$, and ${\bf u}_k \in\mathcal{U} := \{ {\bf u}  \mid -1 \le { u} \le 1  \}$, respectively.
We consider the following quadratic objectives,
 \begin{equation}
 \label{eq:empc_qp}
   \mathcal{L}_{\texttt{MPC}} =  \sum_{k=0}^{N-1}  \big(
 || {\bf x}_k||_{Q_x}^2  + || {\bf u}_k ||_{Q_u}^2 \big).
 \end{equation}
 
\begin{figure*}[htbp!]
    \centering
    %\captionsetup{justification=centering,font=}
    \begin{minipage}{0.46\linewidth}
    \centering
    \includegraphics[width = \linewidth]{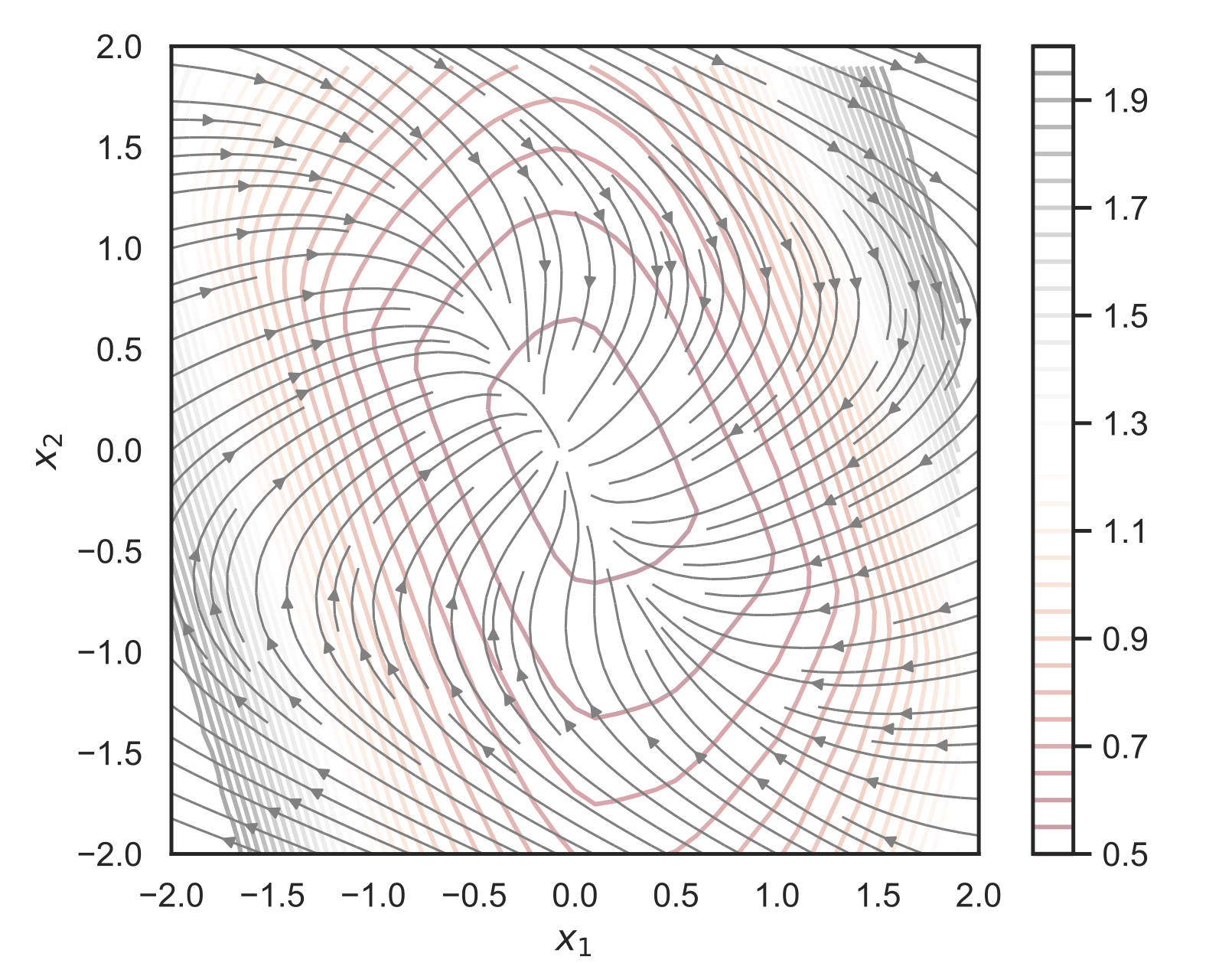}
    \caption{\small{Closed-loop phase portrait with overlaid Lyapunov contours for double integrator.}}
    \label{fig:DI_phase}
    \end{minipage}
        \begin{minipage}{0.2\linewidth}
            \end{minipage}
    \begin{minipage}{0.4\linewidth}
    \includegraphics[width = \linewidth]{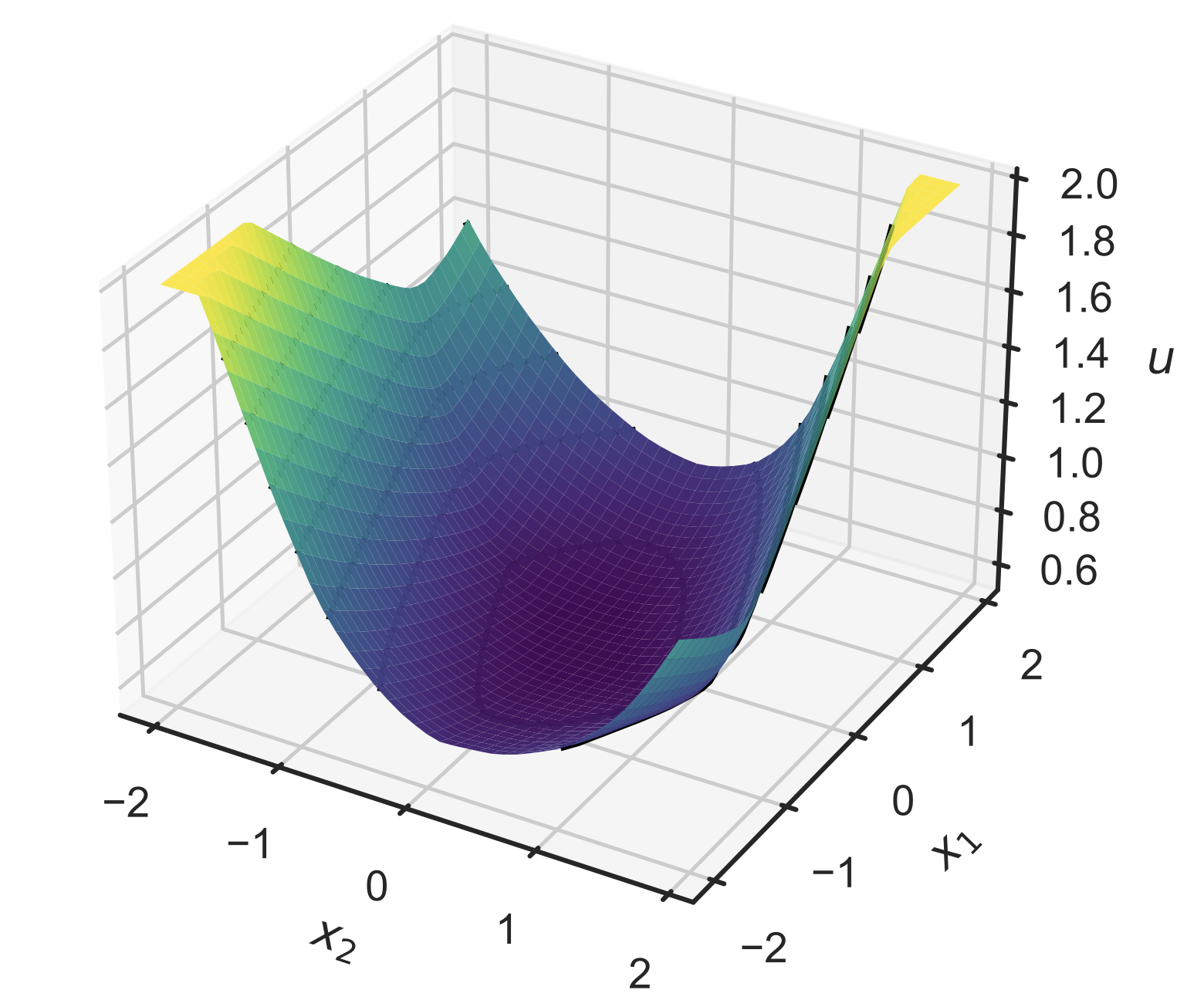}
     \caption{\small{Learned Lyapunov function for double integrator.}}
    \label{fig:DI_lyapunov}
    \end{minipage}
        \begin{minipage}{0.46\linewidth}
        \centering
        \includegraphics[width = \linewidth, height = 4.5 cm]{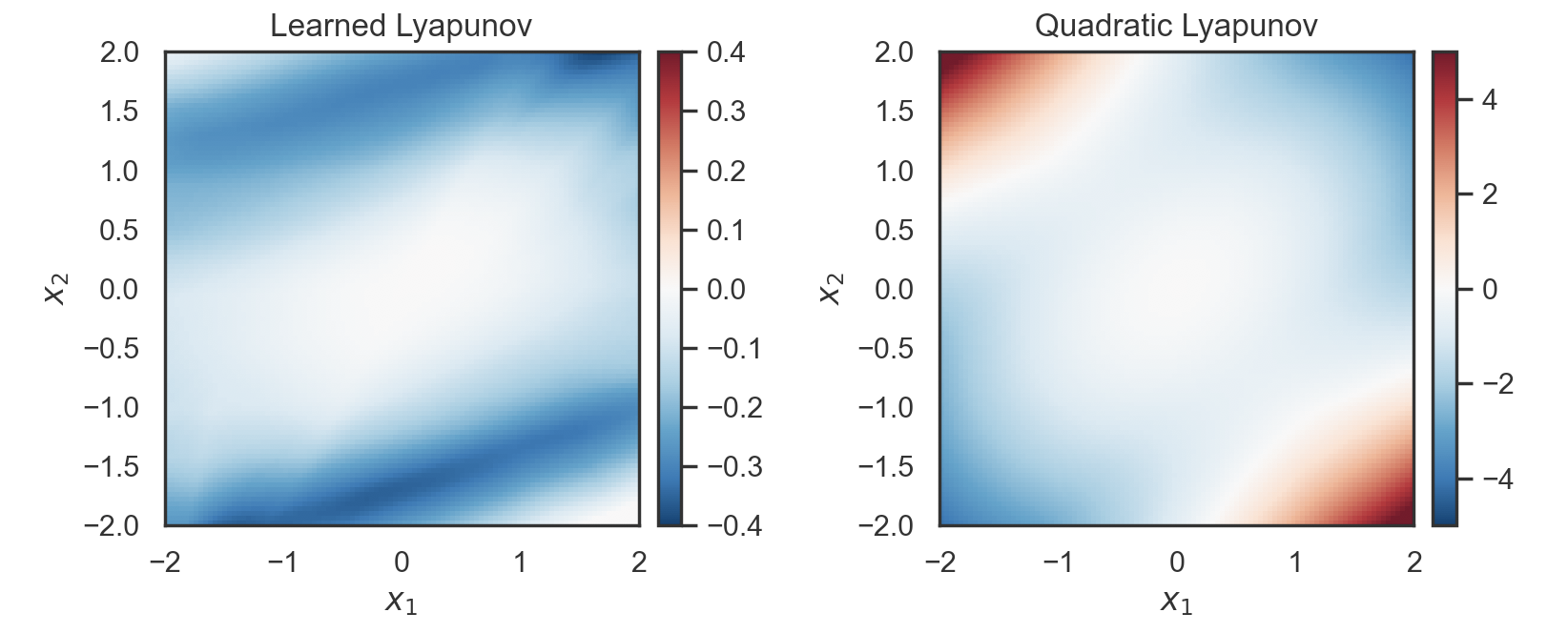} % first figure itself
        \caption{\small{Stability regions via discrete-time Lyapunov stability condition~\eqref{Discrete-time Lyap} for double integrator.}}
        \label{fig:DI_Vdiff}
    \end{minipage}
        \begin{minipage}{0.2\linewidth}
        \end{minipage}
    \begin{minipage}{0.4\linewidth}
        \centering
        \includegraphics[width = \linewidth]{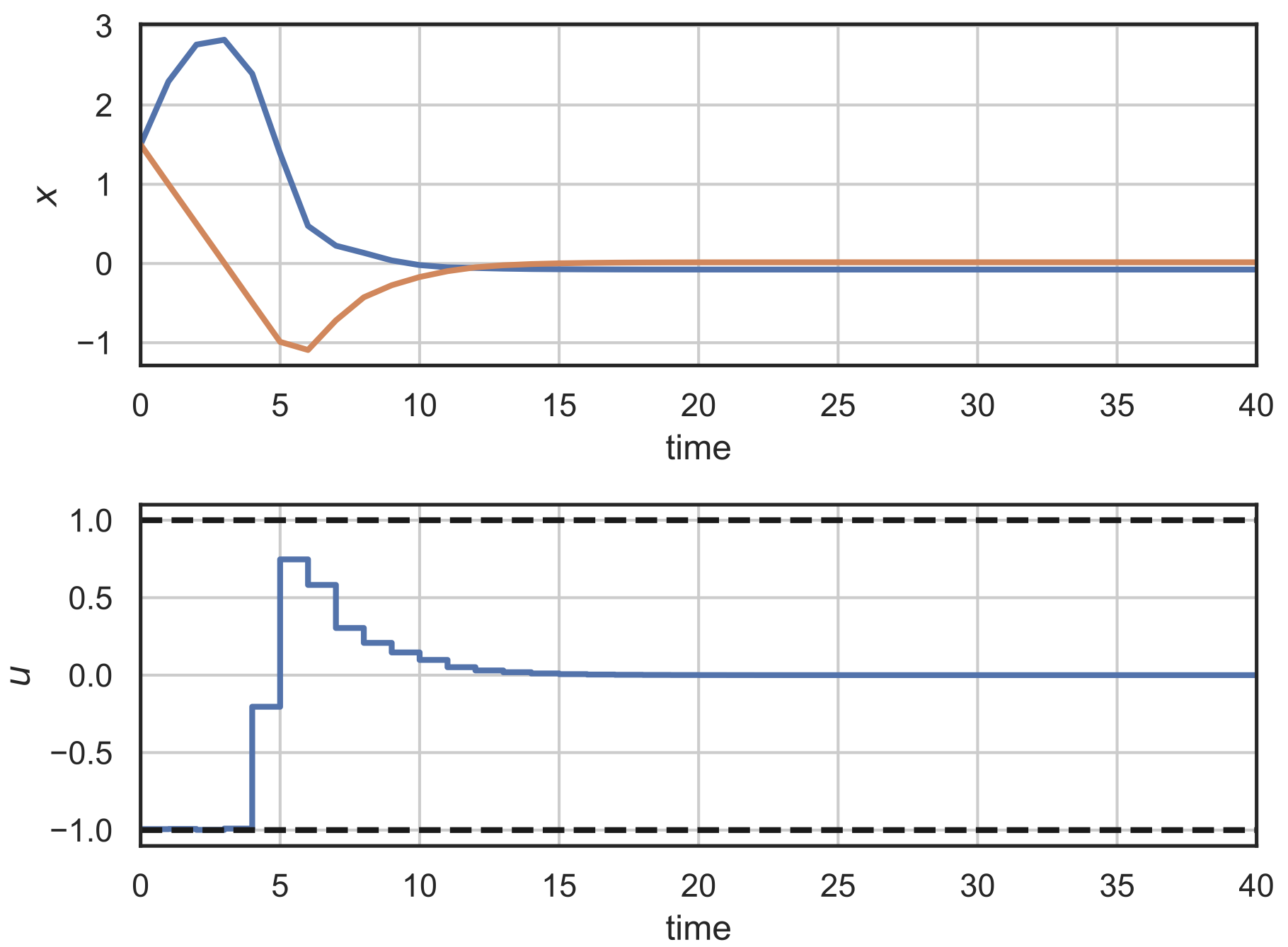} % first figure itself
        \caption{\small{Closed-loop state and control trajectories for double integrator.}}
    \label{fig:DI_time-series}
    \end{minipage}
    \end{figure*}

 For learning neural Lyapunov DPC, we intend to learn a full state feedback neural policy
 $\boldsymbol \pi_{\theta}({\bf x}_k)$
 using Algorithm~\ref{algo:NLDPC}. 
 We also impose 
 terminal set constraints ${\bf x}_N \in \mathcal{X}_f :=  \{ {\bf x} \mid -0.1 \le { x} \le 0.1 \}$. We also impose the Lyapunov constraints given in \eqref{eq:DPC}. The weights are considered to be $Q_V = 2, Q_{h} = 10$, $Q_{g} = 100$, $Q_{\mathcal{X}_f} = 1$, while for the control objective~\eqref{eq:empc_qp} 
 we consider prediction horizon $N = 1$, and weights $Q_x = 5$, $Q_u = 0.5$.
 The neural policy~\eqref{eq:dnn} $ \pi_{\theta}({\bf x}): \mathbb{R}^2 \to \mathbb{R}$ is trained with $4$ layers, $20$ hidden states, with bias, and \texttt{ReLU} activation functions. For the Lyapunov function $V_\phi (\bf x)$ we use $8$ layers with $40$ hidden states and \texttt{ReLU} activations.
 The training set  $\mathcal{X}^{\text{train}}$ consists of $3333$  normally sampled initial conditions ${\bf x}^i_0$  fully covering the admissible set  $\mathcal{X}$. The training has been performed for $300$ epochs. Fig.~\ref{fig:DI_phase} shows the phase portrait of the two states which has been stabilized and the level curves of the learned Lyapunov function has also been superimposed. The learned Lyapunov function is shown in Fig.~\ref{fig:DI_lyapunov} ,and the closed-loop state trajectories in Fig.~\ref{fig:DI_time-series} which shows the closed-loop stabilizing performance of NLDPC. Fig.~\ref{fig:DI_Vdiff} shows the finite differences of the learned Lyapunov function for two consecutive time steps of the closed-loop dynamics in the left panel, and compared with the quadratic Lyapunov ${\bf x}^\top {\bf x}$ in the right panel. The figure shows that the quadratic Lyapunov can give a conservative safe region of operations whereas the learned version using NLDPC is less conservative and tend to justify the actual learned closed-loop behaviour.

 \subsection{Example 2: Stabilizing PVTOL aircraft model}
 
\begin{figure*}[htbp!]
    \centering
    %\captionsetup{justification=centering,font=}
    \begin{minipage}{0.46\linewidth}
    \centering
    \includegraphics[width = \linewidth]{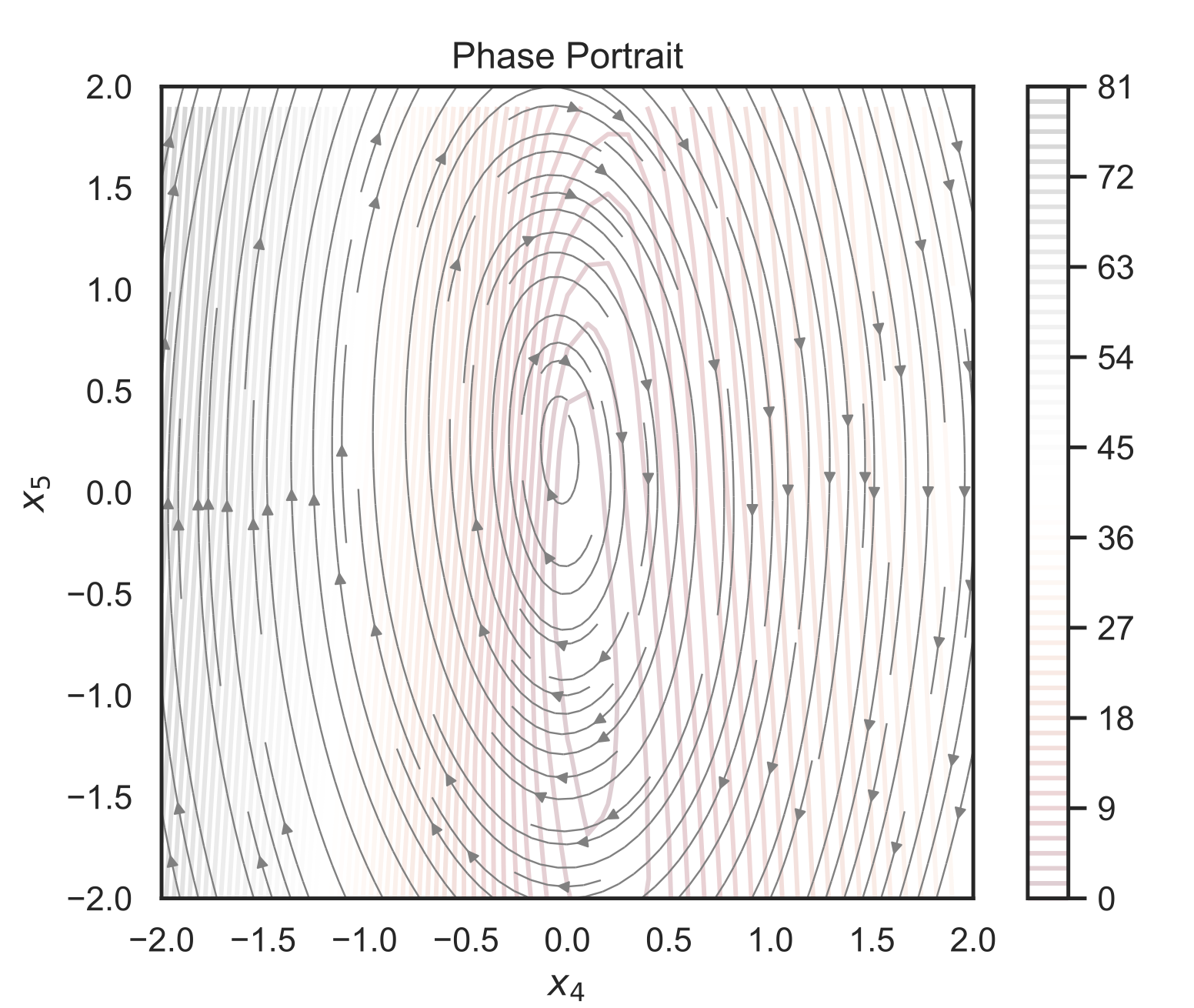}
    \caption{\small{Closed-loop phase portrait with overlaid Lyapunov contours for PVTOL aircraft. Visualized on 2D slice of the 6D state space.}}
    \label{fig:vtol_phase}
    \end{minipage}
        \begin{minipage}{0.2\linewidth}
        \end{minipage}
    \begin{minipage}{0.39\linewidth}
    \includegraphics[width = \linewidth]{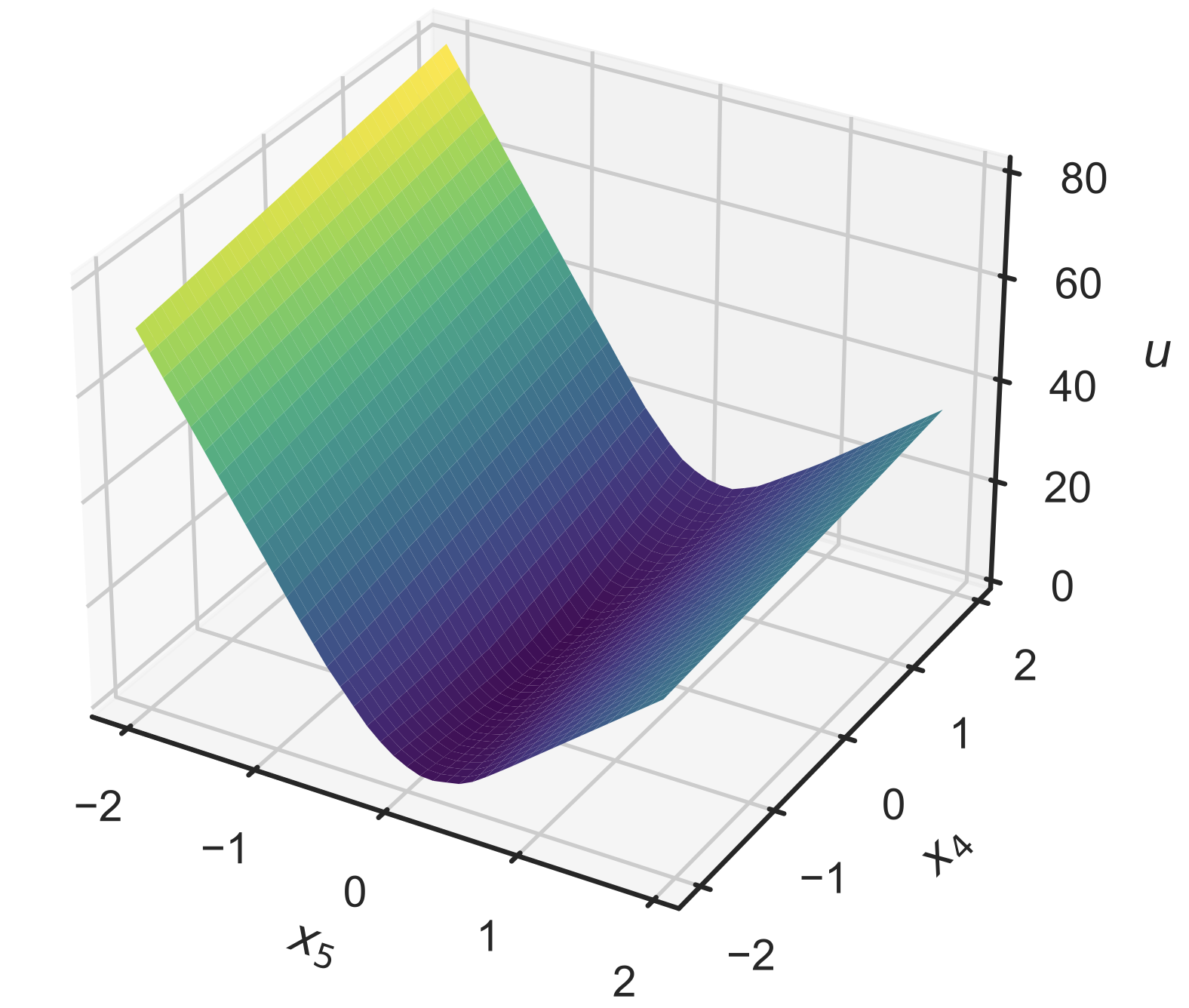}
     \caption{\small{Learned Lyapunov function for PVTOL aircraft. Visualized on 2D slice of the 6D state space.}}
    \label{fig:vtol_lyapunov}
    \end{minipage}
        \begin{minipage}{0.46\linewidth}
        \centering
        \includegraphics[width = \linewidth, height = 4.5 cm]{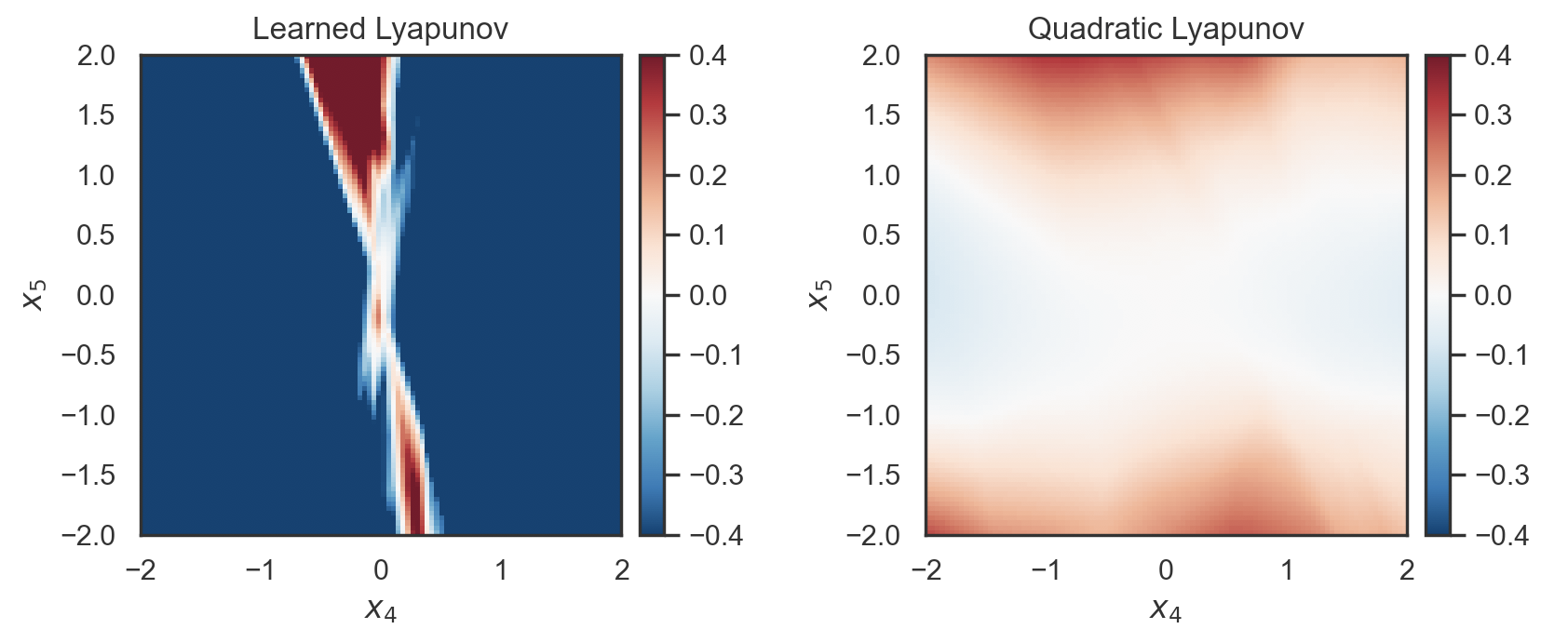} % first figure itself
        \caption{\small{Stability regions via discrete-time Lyapunov stability condition~\eqref{Discrete-time Lyap} for PVTOL aircraft. Visualized on 2D slice of the 6D state space.}}
        \label{fig:vtol_Vdiff}
    \end{minipage}
            \begin{minipage}{0.2\linewidth}
        \end{minipage}
    \begin{minipage}{0.41\linewidth}
        \centering
        \includegraphics[width = \linewidth]{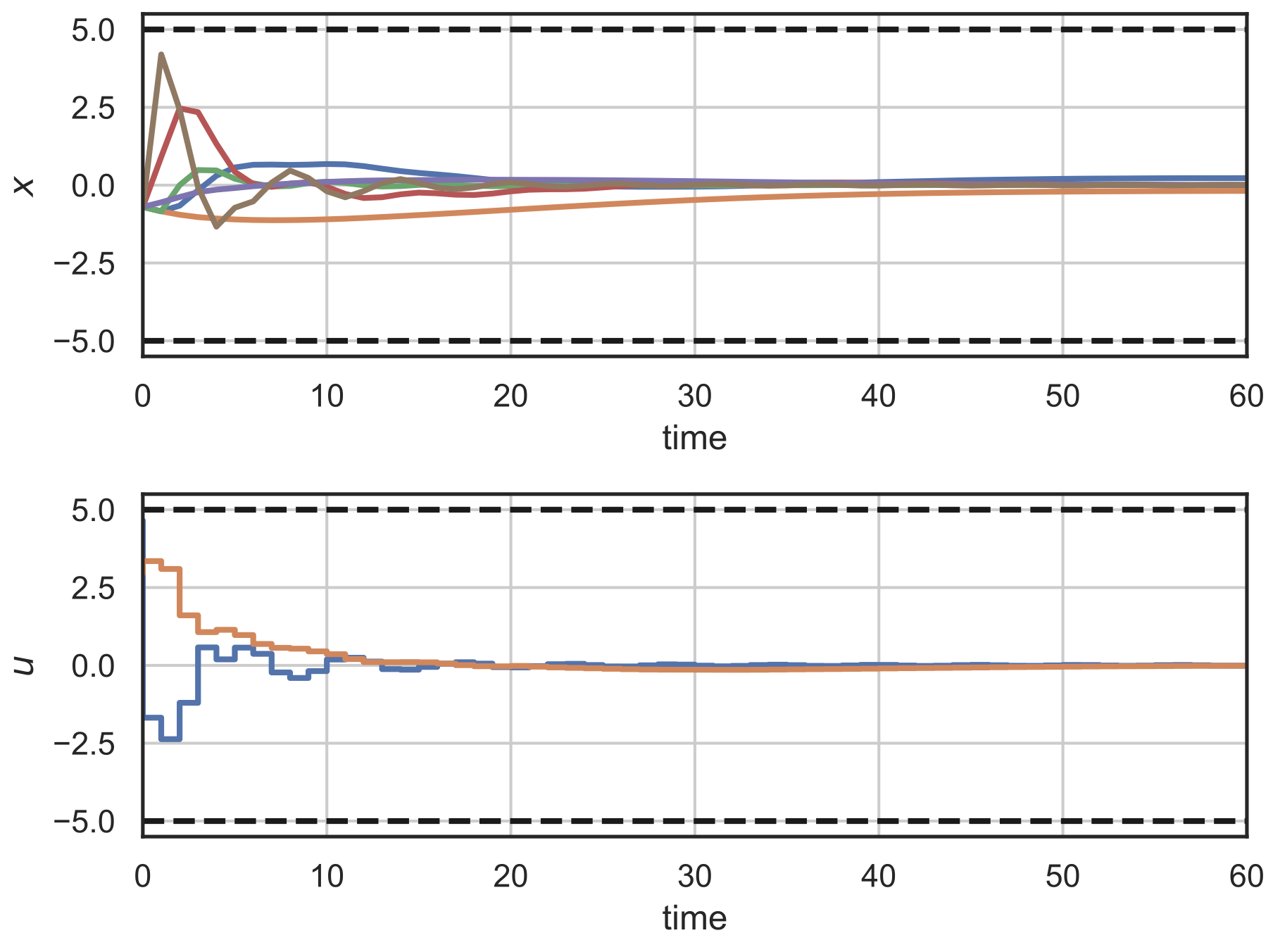} % first figure itself
        \caption{\small{Closed-loop state and control trajectories for PVTOL aircraft.}}
    \label{fig:vtol_time-series}
    \end{minipage}
    \end{figure*}

 For the next example, we considered a planar  vertical take-off and landing (PVTOL) aircraft model as appeared in~\cite{Astromfeedbacksystems}. We have ${\bf x}_k \in \mathbb{R}^6$ representing  positions $(x, y)$,  velocities $(\dot{x}, \dot{y})$, and orientation of the center of mass  ${\theta}$, and its derivative  $\dot{\theta}$. The control inputs ${\bf u} \in \mathbb{R}^2$ are the forces from the aircraft thrusters. We use a discrete-time PVTOL model with sampling time of $0.2$ seconds. Here, the quadratic control objective uses weights $Q_x = 3$, $Q_u = 0.1$ with prediction horizon $N =10$, and 
 subject to state and input constraints
 ${\bf x}_k \in \mathcal{X} := \{ {\bf x} \mid -5 \le { x} \le 5 \}$, and ${\bf u}_k \in\mathcal{U} := \{ {\bf u} \mid -5 \le { u} \le 5 \}$. The penalty weights  are considered to be, $Q_{V} = 3, Q_{h} = 2,$ $Q_{g} = 2$, respectively. The neural policy ~\eqref{eq:dnn} $ \pi_{{\bf W}}({\bf x}): \mathbb{R}^6 \to \mathbb{R}^{N \times 2}$ is similarly considered with $4$ layers, $20$ hidden states, with bias, and \texttt{ReLU} activation functions. For the Lyapunov function $V_\phi (\bf x)$ we use $8$ layers with $40$ hidden states and \texttt{ReLU} activations, as before. We generate $9000$ normally distributed initial conditions with $3000$ each for training, validation and testing sets. Fig.~\ref{fig:vtol_phase} shows the phase portrait sliced only with the aircraft velocities, and the overlaid learned Lyapunov level curves. The learned Lyapunov function with only the above mentioned dimensions are shown in Fig.~\ref{fig:vtol_lyapunov} showing the positive definiteness. Fig.~\ref{fig:vtol_time-series} shows all the states and control actions in the closed-loop which satisfies all the required constraints. Fig.~\ref{fig:vtol_Vdiff} shows the finite differences of the Learned and the quadratic Lyapunov functions for two time-steps of the closed-loop where again the learned Lyapunov function guarantees a larger region of safety than the quadratic one. 

\section{Conclusions}
This paper presented a differentiable programming-based framework for joint learning of explicit neural predictive control policies and neural Lyapunov functions. We bring forth the idea of constraining the parametric optimal program with learnable Lyapunov-based stability certifications. We supplement our training framework with initial condition sampling-based probabilistic guarantees using Hoeffding's inequality. Numerical experiments on standard dynamic system examples of double integrator and PVTOL aircraft model substantiate the algorithmic and theoretical results. Future research will look into more exhaustive experiments with nonlinear system models and using learned identification-based models in the NLDPC framework including different architectures of neural Lyapunov function candidates. 
On the theoretical side, we intend to expand the framework to support time-varying Lyapunov functions and dissipative system conditions used in economic MPC to deal with plant model mismatch and measurement noise.

\section*{Acknowledgement}
This research was supported by the U.S. Department of Energy, through the Office of Advanced Scientific Computing Research's “Data-Driven Decision Control for Complex Systems (DnC2S)” project. Pacific Northwest National Laboratory is operated by Battelle Memorial Institute for the U.S. Department of Energy under Contract No. DE-AC05-76RL01830.

\bibliographystyle{IEEEtran}
\bibliography{ref}

\end{document}